# Evaluation of Wafer-Scale SOT-MRAM for Analog Crossbar Array Applications

Samuel Liu[1,2], Chen-Yu Hu[3], Ming-Yuan Song[3], Xinyu Bao[3], and Jean Anne C. Incorvia[1,2]

[1] Department of Electrical and Computer Engineering, The University of Texas at Austin, Austin, TX, USA

[2] Microelectronics Research Center, The University of Texas at Austin, Austin, TX, USA

[2] Corporate Research, Taiwan Semiconductor Manufacturing Corporation, Hsinchu, Taiwan

**ABSTRACT**

Analog crossbar arrays consisting of emerging memory devices can greatly alleviate the computational strain required by vector matrix multiplications for neural network applications. The ability to produce spin orbit torque-magnetic random-access memory (SOT-MRAM) at wafer-scale positions SOT-MRAM as a strong memory candidate. In this work, we fabricate and measure 300 mm-compatible SOT-MRAM with 150% tunnel magnetoresistance ratio, fast (2 ns) and low voltage (<1 V) operation, low energy dissipation (350 fJ), low write noise (0.1%), and low device-to-device variation of 10%. Through 2-bit quantization aware training and noisy training as mitigation techniques, the measured SOT-MRAM devices attain 95% on MNIST. The bi-stable anisotropy and stochastic switching of SOT-MRAM can additionally be leveraged for stochastic training of binary neural networks, able to reach ideal accuracy for a single device. Lastly, the devices were evaluated on implementation of probabilistic graph modeling and the interplay of tunnel magnetoresistance ratio, probability curve distribution, and conductance noise was shown to reduce potential errors in implementation. Through these results, SOT-MRAM is shown to be a uniquely effective candidate for implementation of crossbar accelerators in memory- and energy-limited applications, able to take advantage of stochastic operation and bi-stability to beneficial results in neural network applications.

**BODY**

As the widespread use of machine learning algorithms demands ever greater computational weight, unconventional computing architectures are becoming increasingly important to address this demand[1].

Among them, analog crossbar array circuits leveraging non-volatile memories have been proposed to accelerate vector matrix multiplications, one of the most costly and ubiquitous mathematical operations in machine learning applications[2,3]. A wide variety of memories have been explored to build these crossbar arrays, ranging from more traditional SRAM[4,5], DRAM[6], and flash[7] memories to less traditional memories such as phase change memory (PCM)[8–11], resistive random access memory (RRAM)[12–16], and magnetic random access memory (MRAM)[17,18], to more experimental memories such as 2D materials[19–22] or ionic material-based electrochemical random access memory (ECRAM)[23–28]. Among these, MRAM has the advantages of relatively fast operation of a few ns via ferromagnets to potential ps-fs via antiferromagnets[29,30], low voltage operation < 1 V, high write endurance up to $10^{15}$ cycles, and the existence of semiconductor processes to produce reliable wafer-scale arrays[31–35]. MRAM is based around the magnetic tunnel junction (MTJ), where two ferromagnetic layers sandwich an insulating tunnel barrier and the device is in a low (high) resistance state when the relative magnetization of the two ferromagnetic layers is parallel (anti-parallel). While previous works have explored implementations of wafer scale spin transfer torque-MRAM (STT-MRAM) in analog crossbar arrays for neuromorphic computing applications[17], spin orbit torque-MRAM (SOT-MRAM) has only recently been applied to wafer scale memory applications[18,33,35,36]. While SOT-MRAM is a three-terminal memory in contrast to STT-MRAM, a two-terminal memory, leading to increased area cost, SOT-driven magnetization switching has advantages of higher speed (down to ps timescales)[37], up to 10 times better energy efficiency[38], and up to 10 times lower switching voltage[38] due to the higher comparative efficiency of SOT switching vs. STT switching (*e.g.* ~0.58 charge-to-spin conversion in CoFeB spin valves[39] vs. 0.9 in Pt-based SOT structures[40]). Additionally, the physical separation of the higher current write channel through the heavy metal SOT layer, and the lower current read channel through the tunnel barrier facilitates potentially higher write endurance. This is because the main cause of device degradation in MTJ-based devices is degradation of the insulating tunnel barrier due to Joule heating[41]. However, for neuromorphic applications of synaptic weight representation, SOT-MRAM retains the same drawback of STT-MRAM where a scaled MTJ can usually only represent two states due to shape, magnetocrystalline, or interfacial anisotropy effects. While much previous research has explored how more states can be introduced to magnetic devices through multi-domain nucleation[42,43], domain walls[44–50], skyrmions[51–54], and magneto-ionics[55–60], we focus on applications that can benefit from the highly nonlinear switching effect induced by magnetic anisotropy. In this work, we measured SOT-MRAM devices sampled from wafer-scale fabrication of 4Kb memory arrays and analyze their projected performance on data-driven applications of neural network inference acceleration on 2-bit quantized networks, stochastic training of binary neural networks, and probabilistic graph model representation. We identify implementation strategies and areas of improvement and establish that SOT-MRAM is well-suited for limited-resource edge computing applications due to high speed, high endurance, high energy efficiency, and distinct switching characteristics derived from magnetic physics. These results show there is an important role in neuromorphic computing for SOT-MRAM non-volatile memories that have stochastic switching dynamics coupled with two stable resultant states.

To evaluate the expected performance of the devices at the state-of-the-art level, a 4Kb array of SOT-MRAM devices were fabricated using the same methodology described in Ref. [18]. Figure 1a is a depiction of the typical structure of the in-plane magnetic anisotropy SOT-MTJ device along with the electrical operation of the bitcell. Figure 1b depicts a scanning electron microscope capture of the cross-section of the devices. The films were grown using sputter deposition followed by 400 °C annealing for 30 minutes. The resultant film had a RA of 10 $\Omega*\mu m^2$ and an average tunnel magnetoresistance (TMR) of 170% as measured by current in-plane tunneling. The heavy metal layer is composite W, which was measured to have a spin Hall angle of 0.6. The devices were then fabricated by patterning with electron-beam lithography followed by etching using a hard mask reactive ion etch followed by an Ar inductively couple plasma etch, which was done to avoid damage to the sidewalls of the MTJ. Lastly, an ion beam etch was performed at low energy to eliminate redeposition on the MTJ sidewalls and preserve the SOT underlayer. The MTJ devices were patterned into ellipses of dimension 75 nm by 230 nm. Figure 1c shows 100 field loops of a single SOT-MTJ device, showing sharp, symmetric switching current loops with high effective TMR at 150% and high thermal stability 152. When operated with a pulse width of 10 ns, the resulting energy dissipation is an average of 350 fJ per write event. As previously shown[18], devices show no degradation at $>10^{12}$ write pulses (followed by read pulses), showing high endurance effective for a wide variety of memory and in-memory computing applications. These characteristics are well-suited for analog neural network inference acceleration applications, where retention and conductance stability have been shown to be two of the more important factors when evaluating inference accelerators[61]. Additionally, low write energy and fast writes are also important as the size of the models can exceed the amount of crossbar memory available, leading to necessary writes to perform matrix multiplication acceleration[62]. While the lack of analog states can be a drawback for inference applications, the bi-stability of MTJs can be leveraged using quantization aware training.

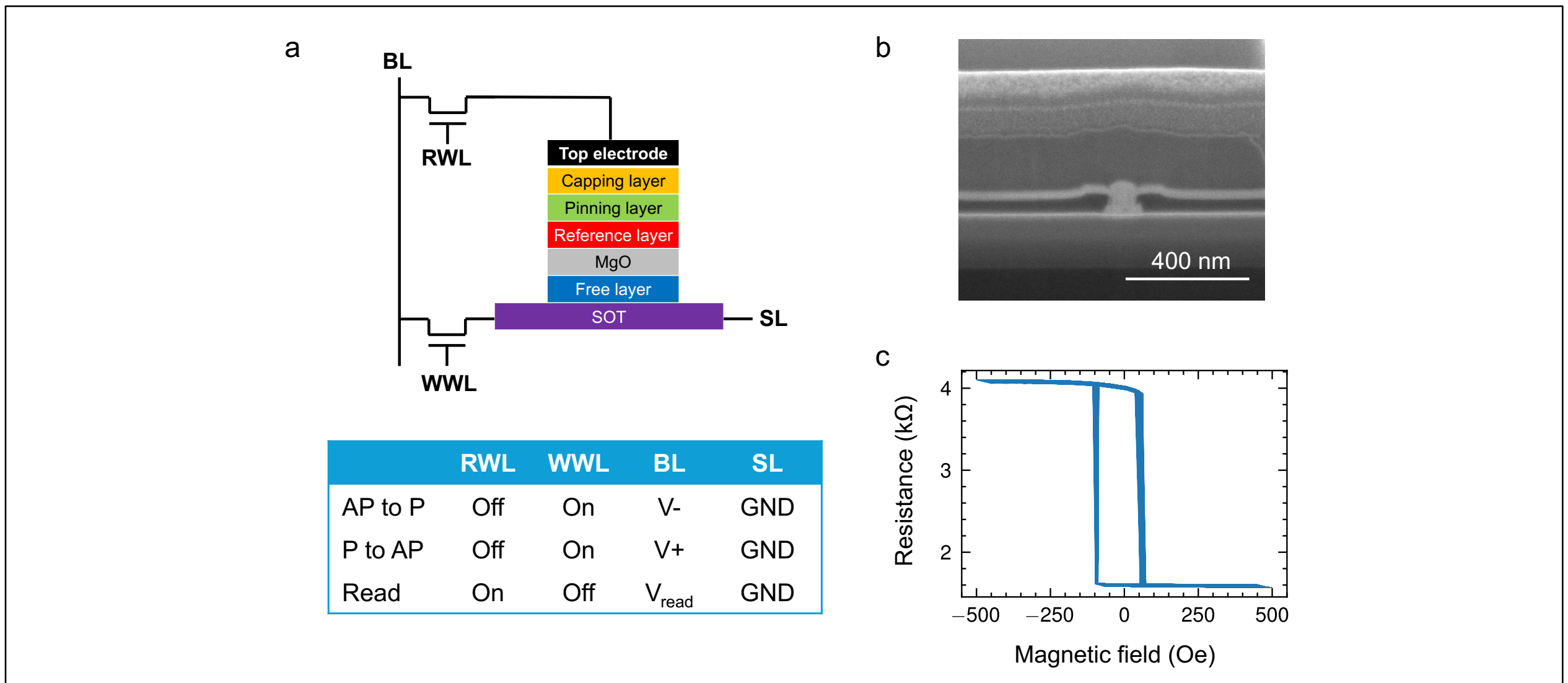

| | RWL | WWL | BL | SL |
|---|---|---|---|---|
| AP to P | Off | On | V- | GND |
| P to AP | Off | On | V+ | GND |
| Read | On | Off | $V_{read}$ | GND |

Figure 1. (a) Schematic of SOT-MRAM 2T-1R bitcell with operational table. The bit line (BL), read word line (RWL), write word line (WWL), and source line (SL) connections to the device are shown. (b) Image taken in focused ion beam of cross-section of SOT-MRAM device. (c) In-plane field switching curves of a single device for 100 cycles.

Due to the stable two states, we propose the use of SOT-MRAM devices for acceleration of 2-bit quantized neural networks, meaning that weights are constrained to values of -1, 0, and 1, represented by two SOT-MRAM devices per cell. While this limits the effective resolution of the crossbar array elements, very little digital memory storage is required to maintain a backup of the weights, synergizing well with the application to resource-limited systems[63]. Networks that are quantized to 2 bits post-training traditionally suffer severe accuracy penalties, but this is alleviated through quantization aware training (QAT)[64]. LeNet-5 architecture convolutional neural networks (CNNs) shown in Fig. 2a were trained using Keras[65] on the MNIST handwritten digits dataset[66] and evaluated in CrossSim[67], a crossbar simulator that samples from measured data, with ideal differential weights at varying quantization level post-training, shown in Fig. 2b. The conventionally trained model (blue curve) shows a noticeable inference accuracy drop when quantized to 3 bits, with a severe drop-off at 2-bit quantization. However, when the CNN is trained using QAT (green curve), this accuracy is largely recovered, with a maximum validation accuracy of 98.2% compared to a maximum validation accuracy in the full precision model (assumed to be operated at 8 bits) at 99.2%, a small drop of ~1% with a 94% reduction in model memory usage compared to 32-bit floating precision.

To predict the performance of the network accurately, the high resistance state (HRS) and low resistance state (LRS) of 100 different devices on the wafer were measured, resulting in the distribution shown in Fig. 2c. The devices have an average TMR of 166%. The spread of the conductance values is approximately proportional, where the HRS has a conductance spread of $k_{\sigma,HRS} = \frac{\sigma_{G,HRS}}{\mu_{G,HRS}} = 9.4\%$ and the LRS has a conductance spread of $k_{\sigma,LRS} = \frac{\sigma_{G,LRS}}{\mu_{G,LRS}} = 10.2\%$, where $\sigma_G$ and $\mu_G$ are respectively the standard deviation and mean of the conductance state. This variation is then applied into CrossSim by generating crossbars with device conductance variations matching the data and the 2-bit QAT CNN is modeled on the array of representative devices. Figure 2d shows the inference accuracy of the array over 20 runs with different seeds (blue), with an average accuracy of 88% with a standard deviation of 3.2%, shown by the shading. This is also compared with an array that is evaluated with only the write noise of a single device (green), with an average accuracy of 95.1% and a standard deviation of 0.2%. The write noise of a single device, at 0.107%, was calculated by analyzing the HRS and LRS of the device in Fig. 1c at 0 field. This is a projection of the limit of the maximum accuracy achievable with a SOT-MRAM array if device-to-device variation was minimized. The black series depicts a generic SOT-MRAM device with weight-proportional noise. This data is meant to represent wafer-scale SOT-MRAM of various device-to-device variations. The intercept of the weight-proportional devices and the SOT-MRAM array is at $k_\sigma$ = 10.4%, where $k_\sigma$ is the weight proportional noise level, showing agreement between the experimental data and the weight-proportional model. From the results in Fig. 2d, the device-to-device variation results in a significant reduction in inference accuracy compared to the single device case.

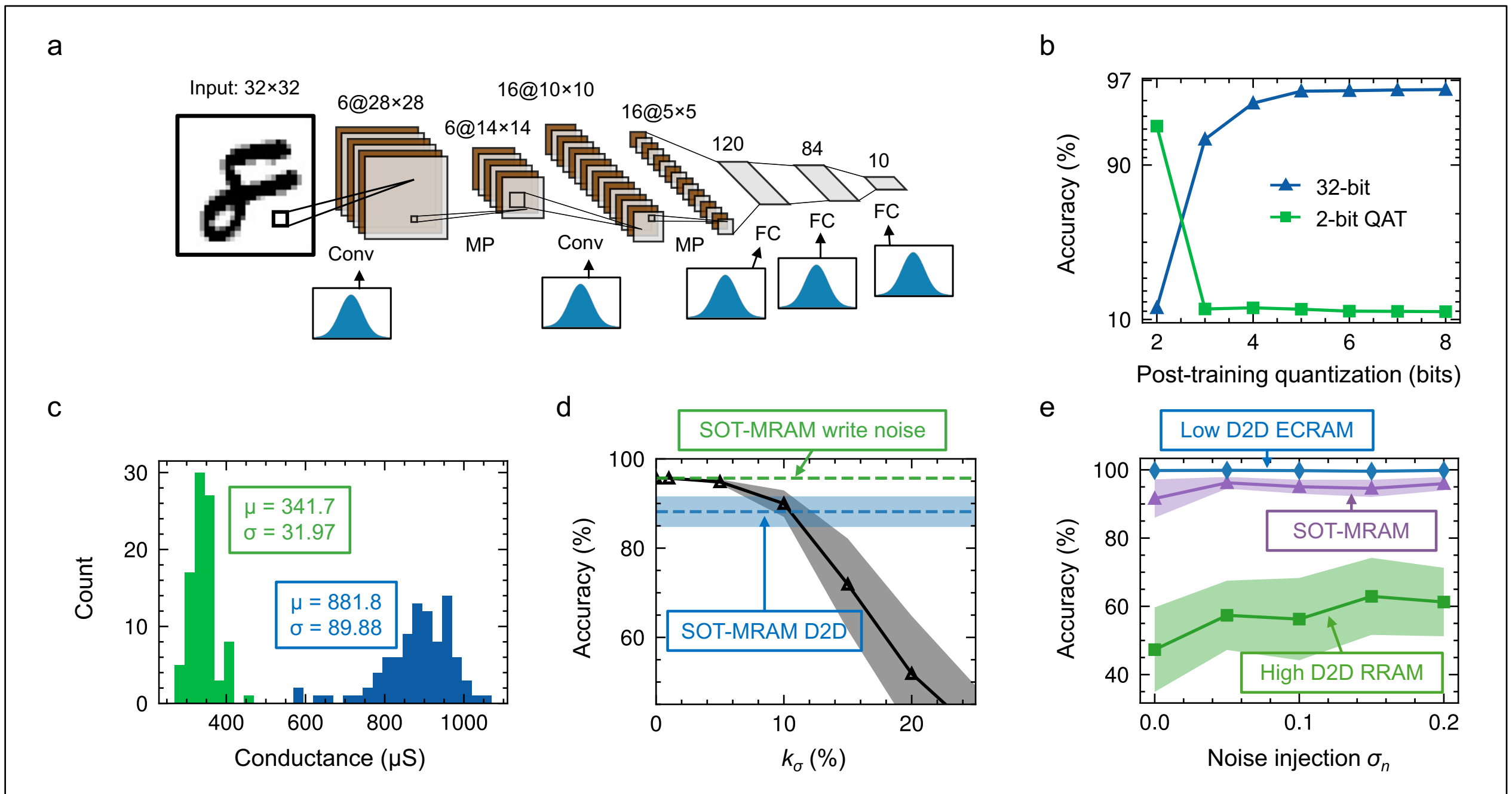

Fig 2. (a) LeNet-5 architecture with gaussian noise injection in weight layers. (b) Inference accuracy of normally trained network and 2-bit quantization aware network performed at varying quantization levels. (c) Measured conductance spread of 100 devices. (d) Crossbar array inference accuracy using measured device-to-device variation (blue) compared to accuracy using SOT-MRAM write noise of measured single device (green) and modeled proportional noise model (black). (e) Inference accuracy of SOT-MRAM device (purple) compared to wafer-scale $TaO_x$ (green) and electrochemical memory (blue) at varying noise injection sigma.

While tightening the device-to-device variation is necessary to increase the inference performance, at variations of approximately 5%, there is still a noticeable accuracy drop. One strategy to alleviate this by using noise injection during training, which is applied in each weight layer as shown in Fig. 2a. Noise injection has been shown to aid convergence during training to better error minima[68] and result in higher noise tolerance during inference[69,70]. Figure 2e shows inference accuracy for varying training noise injection levels (arb. units) for the characterized SOT-MRAM devices along with an example device with relatively high write noise and device-to-device variation at 23.1% ($TaO_x$ RRAM[71]) and example relatively low write noise and device-to-device variation at 1.1% (ECRAM[23]) for comparison. The results indicate that at moderate noise injection levels, the performance of the SOT-MRAM array increases by 6% at the maximum improvement, matching expectations that the network becomes more noise resilient. However, this trend is not as clear with the characterized RRAM devices with significantly worse variation, indicating that the variation of the SOT-MRAM devices falls within an accepted variation where noise injection can result in increased inference accuracy for the system. Supplementary Fig. S1 shows the validation accuracy of the trained networks, showing that the noise applied to the network training eventually becomes large enough to prevent the network from reaching a better error minimum, resulting in worse accuracy. The interplay between getting a better error minimum and training the network to be noise resilient results in an effective sweet spot for noise injection during training, that the measured SOT-MRAM array meets.

Due to the measured high speed and high endurance of the SOT-MRAM devices, online neural network training acceleration is a promising application for the devices. The binary weight nature of the SOT-MRAM can be leveraged to accelerate binary neural networks (BNNs), shown in Fig. 3a, which constrain weight and activation values to only the two values of -1 and +1, in contrast to the 2-bit quantized network shown previously in Fig. 2 with -1, 0, and +1 as weights. The architecture of this network was chosen to be a multilayer perceptron (MLP), with 784 input units, 10 output units, and 2 hidden layers of

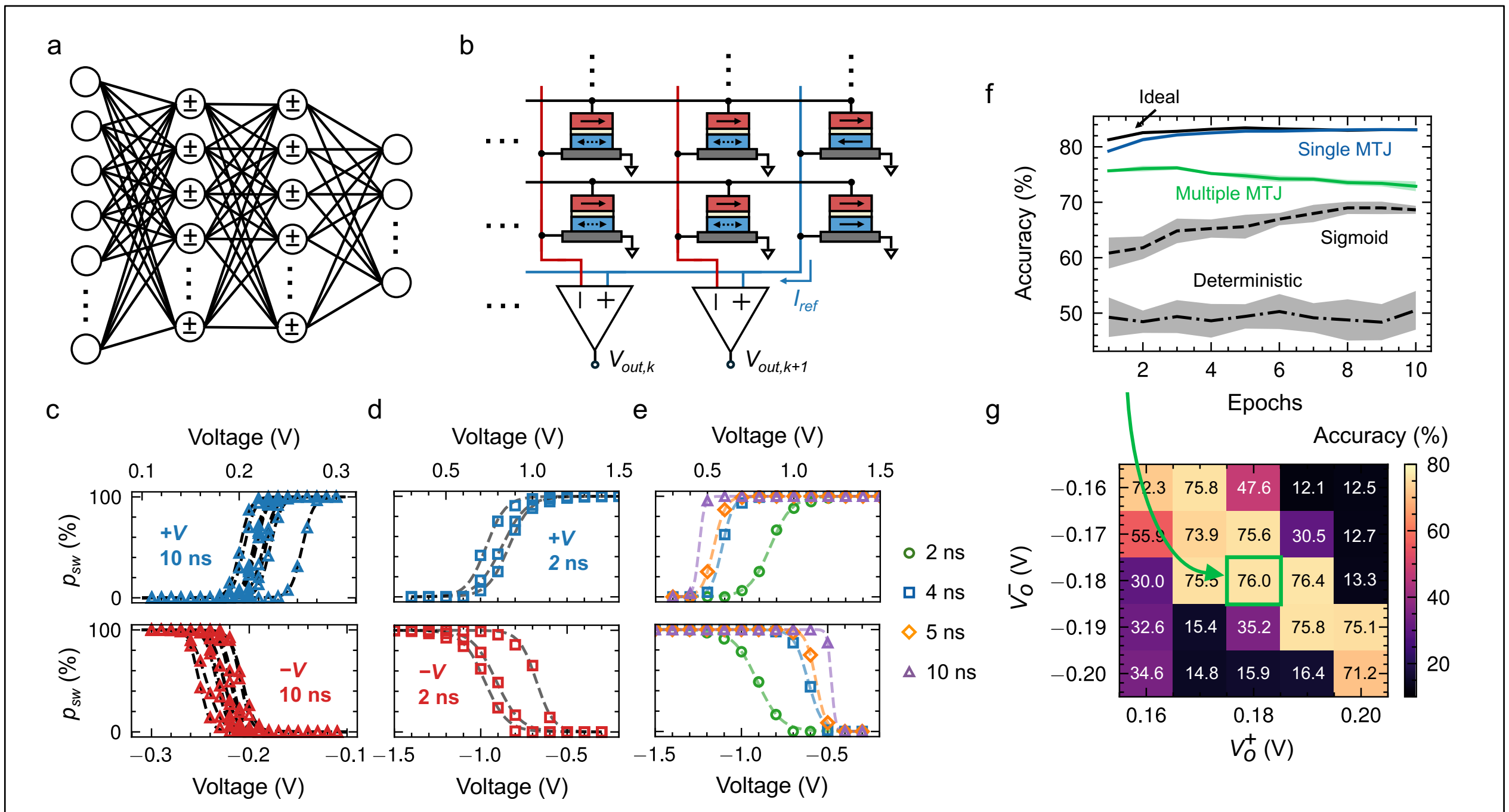

Figure 3. (a) Binarized multilayer perceptron neural network architecture. (b) Crossbar array implementation of binarized neural network, where the column outputs (red) are compared with a column of calibrated devices set to output the zero value (blue) through the reference current $I_{ref}$. Probabilistic switching curves of a set of (c) Measured 10 SOT-MRAM devices using 10 ns pulses and (d) 3 SOT-MRAM devices using 3 ns pulses at varying input voltage. Sigmoidal fittings to the data are depicted in dashed black and grey lines. (e) Measured switching probability as a function of voltages at 2 ns (green circles), 4 ns (blue squares), 5 ns (orange diamonds) and 10 ns (purple triangles). (f) Training validation accuracy of single MRAM devices (blue), MRAM devices with device-to-device variation (green), sigmoid (dashed black) update probability, deterministic (dot-dash black) update probability, and ideal stochastic update (solid black). (g) Training validation accuracy of multiple MRAM heat map of various base positive and negative update voltages.

200 units each. As shown in Fig. 3b, this architecture can be implemented in the crossbar array with only a single device per weight, which is then subtracted from a reference column. The output of the array is also simplified, allowing a single comparator to be used to determine the activation value as opposed to an analog-to-digital converter, often the costliest component in a crossbar array in terms of energy dissipation[72]. Training of this network was implemented in PyTorch[73] by modifying the Adam optimizer[74] with the lookup tables for 10 ns pulse duration shown in Fig. 3c. The training dataset was chosen to be Fashion-MNIST[75], split into 60,000 training images and 10,000 validation images. Though we perform the simulations with the 10 ns lookup tables due to having the highest number of measured devices, measurements for a shorter pulse duration of 2 ns are shown in Fig. 3d. Here, the voltage window for probabilistic switching is much wider than the case for 10 ns, at approximately 0.5 V between 0% and 100% compared to 0.05 V for the 10 ns case, indicating that faster sampling is preferable to reduce the precision requirements of control circuitry. This is corroborated by Fig. 3e, where probability curves of varying pulse duration for a single device are shown.

Figure 3f shows the stochastic training performance in terms of predicted validation accuracy of the array of measured SOT-MRAM devices. When operating using a single MTJ, the network reaches 83% accuracy on average (blue curve), which is close to ideal for this network and task (black curve). When accounting for the device-to-device spread in threshold voltage shown in Fig. 3c, there is a noticeable drop in validation accuracy, maximizing at 76% on average (multiple MTJ green curve). The results are compared with the use of a deterministic update rule and a sigmoid update rule, both of which train worse than the SOT-MRAM network accounting for device-to-device variations. The deterministic update is calculated by taking the sign of the updated weight. The sigmoidal update is obtained by mapping the weights onto the probabilistic function:

$$p = \frac{1}{1 + e^{-\frac{w}{\sigma_s}}}$$

where the probability of switching $p$ is dependent on the post-update weight level $w$ and the width of the sigmoid $\sigma_s$. The deterministic update performs worse because small updates are immediately lost because of the rounding effect of taking the sign of the updated weight. As a result, only a sufficiently large update can change the weight, leading to quick saturation of learning. For the sigmoidal update, there is a small chance for weights at the extreme values of -1 and +1 to flip to the opposite value. This leads to an effective ambient noise during training that is detrimental to accuracy. This contrasts with the SOT-MRAM devices, which can only have a chance to switch to +1 (-1) during positive (negative) updates, leading to behaviors that are more similar to backpropagation-based learning rules. The ideal stochastic update follows these same characteristics and is described in Ref. [76]. While updates to other types of stochastic devices follow the same principle, the inherent variability of SOT-MRAM devices remain low because of magnetic anisotropy, which always forces the resistance state of the device to be either high or low, comparable to the read noise of the device, unlike other devices such as RRAM or PCRAM, which can have resistances in the intermediate states. Additionally, since conductance drift in SOT-MRAM devices is solely due to degradation of the tunnel barrier, SOT-MRAM can achieve greater stability and predictability over time and cycles compared to RRAM and PCRAM, where the devices experience drift due to a host of causes such as ion diffusion, grain boundary differences, and generally high stress switching[9,13,14].

While the biases for each device of the SOT-MRAM array could be saved to prevent device-to-device variation impact by forcing all the probability curves to lie up, using the same biasing voltage for all devices is much more simplistic to implement. Figure 3g shows an evaluation of validation accuracy using different bias voltage values for the positive and negative updates for the measured distributions shown in Fig. 3c. The network performs the best when the biasing voltages are above ±0.17 V. This is likely because if the biasing voltage is too low, then small updates would result in no change in the weight, a detriment to the network performance similar to that of the deterministic update. When the bias voltages are ±0.20 V, there is also a loss in accuracy. This is because for a subset of devices, this bias voltage is enough to induce a large chance of switching the device at every update voltage. The result is similar to using a sigmoidal update, where there is an introduction of extra update noise that is not beneficial for training. The low accuracies in the lower left and upper right quadrants of the graph can be explained by the impact of asymmetry on neural network performance, where an asymmetric update severely reduces the performance of networks[49,77]. From the results, we can conclude that while using SOT-MRAM devices to effectively implement crossbar arrays for binary neural network applications is feasible, the biasing of the update rules for the weights must be carefully chosen to preserve the accuracy of the implemented network.

Due to the tunability of the probabilistic switching characteristics of MTJs, various types of stochastic MTJs from superparamagnetic tunnel junctions[78–80] to probabilistic-write STT-MRAM[81,82] have been characterized and engineered to accelerate Ising and simulated annealing algorithms to solve optimization problems[83–89]. Most of these approaches involve a single device used to generate a random bitstream and peripheral circuitry to process the analog signal. Here, the characterization of wafer-scale

SOT-MRAM allows a prediction of performance on applications involving many devices. One such example is the use of many small crossbar arrays to accelerate probabilistic graph model (PGM) calculations. PGMs can be applied to optimization problems such as steady-state estimation[90,91] and the PageRank algorithm[92], widely used for determining web-page relatedness[93]. The relatedness of two webpage nodes can be represented by the proportion of pass vs. no pass that is sampled at the output. A simple PGM with three nodes 0, 1, and 2 is shown in Fig. 4a. The directional weights of the PGM describe a strength of connection that is represented by probability. Here, there are two possible paths for Node 0 to be connected to Node 2. There is a direct connection (orange) with a probability of 0.2 and an indirect connection passing through Node 1 (purple) with probabilities of 0.6 and 0.5. The crossbar array implementation of the graph is shown in Fig. 4b, where the two possible paths for node connection are shown (left, middle), along with the case where neither path is on (right). These small crossbars are sampled $N$ times until a satisfactory distribution can be attained. In practice, many small crossbars can be fabricated to perform the sampling in parallel. Here, the device crossbars were generated and analyzed through Ngspice[94] through the PySpice package[95]. The results of sampling these crossbars constructed from the measured device probability curves (p-curves) shown in Fig. 3c are shown in Fig. 4c. To sample multiple devices, the average of all p-curves was used to construct a lookup table to identify the necessary voltages to attain the targeted probability. The left graph of Fig. 4c depicts the outputs while only using the measured write noise of a single device, while the right graph depicts the outputs with the full device-to-device variation of the conductance. A threshold voltage to determine whether a sample is a pass or a no-pass is shown in the dashed black line, at 11.5 μA, and the input voltage for sampling was set to be 10 mV.

From the right graph of Fig. 4c, it is evident that several of the samples cross the threshold, resulting in an unexpected reading compared to the actual configuration of the crossbar. This ratio of number of expectation errors to the total number of samples is described as $\epsilon_{exp}$. The window where a threshold can be set is small due to the limited on-off ratio of the devices coupled with the device-to-device variation. In Fig. 4d, the same sampling was done again with a projection of the TMR to an idealized value of 600%. Here, even when accounting for the measured device-to-device variations of the wafer, there is a sufficient output current window to set a solid threshold. Figure 4e compares the $\epsilon_{exp}$ for the measured 150% TMR devices and the projected 600% TMR at varying levels of device-to-device variation $k_{\sigma}$. The results indicate that $\epsilon_{exp}$ remains low and rises much more slowly for the 600% TMR devices compared to the 150% TMR devices, indicating that a design metric of higher on/off ratio can mitigate errors introduced from device-to-device variation of the conductance, which becomes more important as the complexity and dimension of the PGM increases.

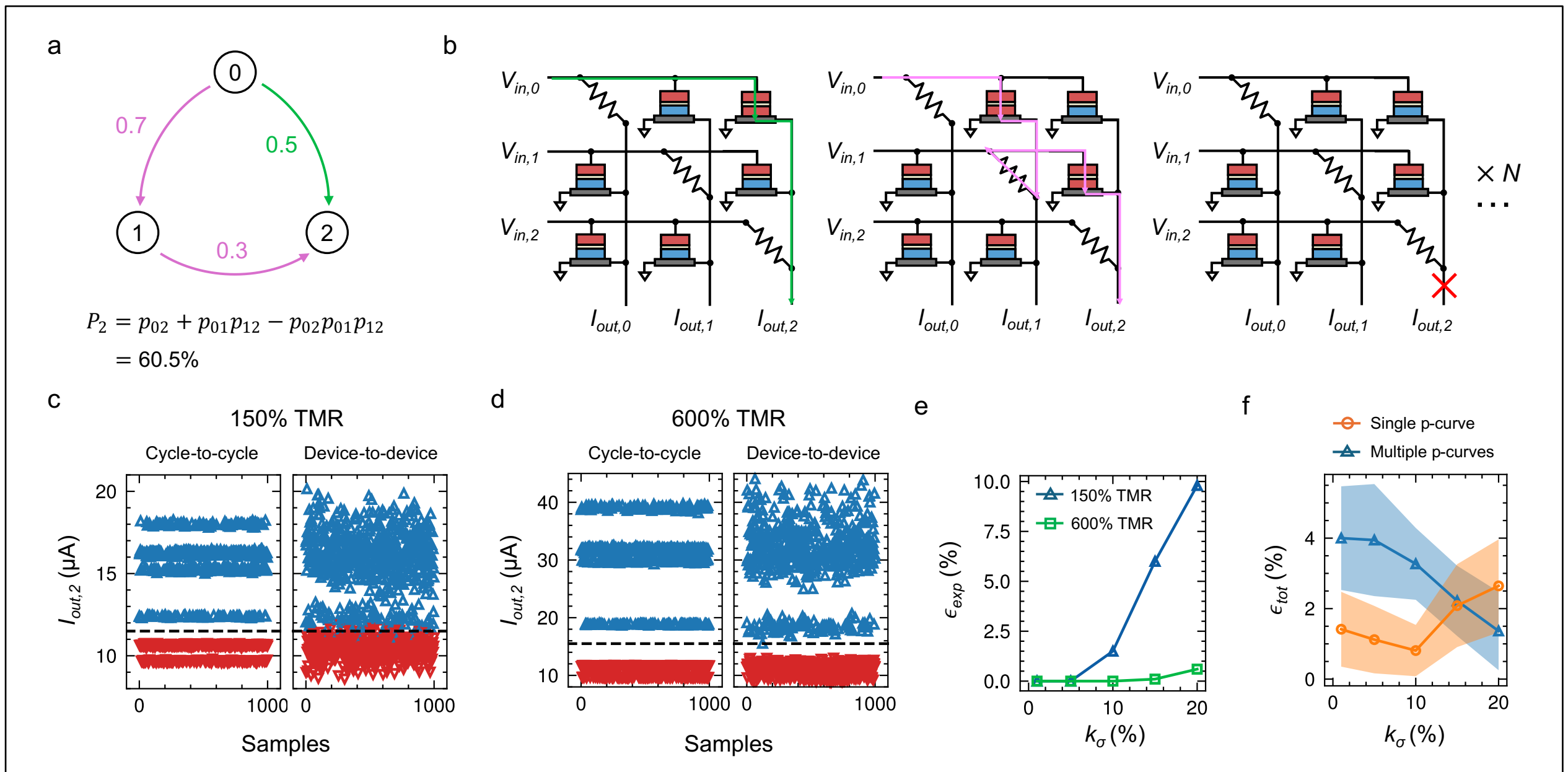

Figure 4. (a) Probabilistic graph model (PGM) where there are two possible paths from 0 to 2. (b) Example crossbar array representations of the PGM. The two possible paths are depicted in green and pink. (c) The current output of column 2 of 1000 different crossbar arrays accounting for cycle-to-cycle (left) and device-to-device (right) variations. This is done for devices with the experimentally measured TMR of 150% (c) and an ideal TMR of 600% (d). (e) Error of output from crossbar configuration $\epsilon_{exp}$ as a function of conductance noise $k_{\sigma}$ for measured TMR of 150% (blue triangles) and ideal TMR of 600%. (f) Total error $\epsilon_{tot}$ as a function of conductance noise $k_{\sigma}$ for networks composing of a single probability curve (orange circles) or multiple probability curves matching wafer-scale variation (blue triangles).

While the error of expectation $\epsilon_{exp}$ might be large because of many samples crossing the threshold due to device-to-device variation, if the threshold is set appropriately between the lowest pass level and highest no-pass level, the statistical result of the actual error can be close to 0% if the distribution of false pass results and false no-pass results is roughly equal. Here, we define the total error to be $\epsilon_{tot} = |P_{2,out} - P_{2,ideal}|$ where $P_{2,out}$ is the sampled probability of a pass and $P_{2,ideal}$ is the ideal probability of a pass. As seen in Fig. 3c, the measured devices have a significant spread in the threshold voltages for the p-curves. In Fig. 4f, the difference in $\epsilon_{tot}$ between wafer-scale device-to-device variation in p-curve spread (blue triangles) and the ideal situation of having a single sampled p-curve (orange squares) is analyzed as a function of device-to-device conductance noise $k_\sigma$. At $k_\sigma$ under 15%, the error matches intuitive results, where the PGMs sampling from a single p-curve performs with lower error than the PGMs sampled from all the measured p-curves. Of note however, at $k_\sigma$ equal to and greater than 15%, this trend reverses. This is most likely because the multiple device lookup table for switching voltages was constructed from the average of all devices, leading to a skewed result for the output at low noise. This is not the case for the single p-curve PGMs, where the lookup table was calibrated perfectly. As a result of the accurate calibration of the single p-curve, deviations due to conductance noise $k_\sigma$ increase $\epsilon_{tot}$. In contrast, for the multiple p-curve PGMs, the increase conductance noise $k_\sigma$ smears the error function for the output probability, in this specific case leading to an increased overlap between the expected output probability $P_{2,ideal}$ and the output probability of the PGMs $P_{2,out}$, a beneficial outcome of the interplay between the error introduced by threshold voltage noise and conductance noise. These results show the most important improvement area for SOT-MRAM for PGM representation is to increase TMR, leading to a larger threshold window and larger simulated PGMs, while errors resulting from device-to-device variation can be mitigated. The interplay between threshold voltage variations and conductance variations can be engineered to reduce errors for PGM sampling of non-ideal devices.

| Device | Speed | Energy | Endurance | On/off ratio | Write noise | Number of states | Device-to-device |
|---|---|---|---|---|---|---|---|
| SOT-MRAM (this work) | 2 ns | 350 fJ | $>10^{12}$ | 2.5 | 0.107% | 2 states | ~10% |
| STT-MRAM[17] | 90 ns | 27 pJ (cell) | $>10^{8}$ | 2 | Not available | 2 states | Not available |
| RRAM[71] | 10 ns | 100 pJ | Not available | 13 | ~11.5% | 8 states | ~23.1% |
| RRAM[16] | 5 ms | 787.5 pJ | $>10^{5}$ | 7 | ~1% | 16 states | Not available |
| RRAM[15] | 80 µs | 400 pJ | $>10^{5}$ | ~10 | ~1% | 16 states | ~29% |
| ECRAM[23] | 100 ns | 3.5 µJ | $>10^{7}$ | 5 | <1% | ~90 states | ~1.1% |
| PCRAM[11] | 50 ns | 1 pJ | $>10^{11}$ | ~100 | ~5% | 12 states | ~9% |
| PCRAM[8] | <100 ns | ~2 nJ | Not available | ~100 | ~8% | ~12 states | Not available |

Table 1. Comparison of synaptic devices with wafer-scale production ability.

Overall, the advantages of the SOT-MRAM in this work are fast speed, low energy dissipation, and high endurance compared to other memories at similar maturity, shown in Table 1. The applications analyzed in this work show how the application can play to those strengths. Memory-efficient 2-bit neural networks with QAT mitigate limited resistance states and device-to-device variation, and binary neural networks and PGM representation leverage the advantages of magnetic anisotropy and probabilistic switching. Additionally, the demonstrated high endurance and low write energy dissipation allowing flexibility even if many write cycles are necessary.

In conclusion, we have shown that SOT-MRAM is a stand-out memory for edge computing applications. Measurements of wafer-scale fabricated SOT-MRAM devices from a 4Kb memory array are used to evaluate SOT-MRAM's effectiveness on several edge-specific applications, including 2-bit quantized inference, stochastic training of binary neural networks, and PGM representation. We identify implementation strategies for the outlined applications and identify increased TMR and reduced device-to-device variation as avenues of improvement for SOT-MRAM, but through comparisons with other technologies such as RRAM and ECRAM, show that SOT-MRAM is a leading candidate for edge computing.

## Methods

The SOT-MRAM devices were fabricated by patterning with electron-beam lithography followed by etching using a hard mask reactive ion etch followed by an Ar inductively couple plasma etch, which was done to avoid damage to the sidewalls of the MTJ. Lastly, an ion beam etch was performed at low energy to eliminate redeposition on the MTJ sidewalls and preserve the SOT underlayer.